\documentclass{appolb}
\usepackage{graphicx}
\usepackage{amssymb,amsmath,bm}
\usepackage{hyperref}

\begin{document}

\thispagestyle{empty}

\title{Search for $\eta$-bound Nuclei}

\author{%
      H. Machner\footnote{h.machner@fz-juelich.de}
\address{%
Institut f\"{u}r Kernphysik, Forschungszentrum J\"{u}lich, 52428 J\"{u}lich, Germany\\
and Fakult\"{a}t f\"{u}r Physik, Universit\"{a}t Duisburg-Essen, Lotharstr. 1, 47048 Duisburg}}
\maketitle

\begin{abstract}
The $\eta$ meson can be bound to atomic nuclei. Experimental search is discussed in the form of final state interaction for the reactions $dp\to{^3\text{He}}\eta$ and $dd\to{^4\text{He}}\eta$. For the latter case tensor polarized deuterons were used in order to extract the s-wave strength. For both reactions complex scattering lengths are deduced: $a_{{}\rm{^3He}\eta }  = \left[ { \pm \left( {10.7 \pm 0.8_{ - 0.5}^{ + 0.1} } \right) + i \cdot \left( {1.5 \pm 2.6_{ - 0.9}^{ + 1.0} } \right)} \right]$ fm and $a_{^4\text{He}\eta }  = \left[ { \pm \left( {3.1 \pm 0.5} \right) + i \cdot \left( {0 \pm 0.5} \right)} \right]$ fm. In a two-nucleon transfer reaction under quasi-free conditions, $p^{27}\text{Al}\to {^3\text{He}}X$, was investigated. The system $X$ can be the bound $^{25}\text{Mg}\otimes\eta$ at rest. When a possible decay of an intermediate $N^*/1535)$ is required, a highly significant bump shows up in the missing mass spectrum. The data give for a bound state a binding energy of 13.3$\pm$1.6 MeV and a width of $\sigma$=4.4$\pm$1.3 MeV.
\end{abstract}


\section{Introduction}
In contrast to the pion-nucleon interaction,
the $\eta$-nucleon interaction at small momenta is attractive and
rather strong. This attraction can be seen from the fact that
the $\eta$ threshold (1488~MeV) is situated just below the
$N^*(1535)$ resonance which couples strongly to the $\eta -N$
channel. Initial calculations by Bhalerao and Liu\cite{Bhalerao85}
obtained attractive s-wave $\eta -N$ scattering lengths. With these values, Haider and Liu\cite{Haider_Liu86} have shown that $\eta$ can be bound in nuclei with  A $\ge$ 10. Other groups have also found similar results
\cite{Garcia-Recio02, Hayano99, Tsushima00}.

On the experimental side results are meager. The first experiments searching for
$\eta$-mesic nuclei at BNL\cite{Chrien88} and LAMPF\cite{Lieb88}
by using a missing-mass technique in the ($\pi^+, p$) reaction reached negative or inconclusive results. Later
 it became clear that the peaks are not necessarily
narrow and that a better strategy of searching for $\eta$-nuclei is
required. Furthermore, the BNL experiment was in a region far from
the recoilless kinematics, so that the cross section is
substantially reduced\cite{Hirenzaki07}. More
recently, the existence of $\eta$-mesic ${\rm{^3He}}$ was claimed to have
been observed in the reaction $\gamma{^{3}\text{He}}\to \pi^0pX$ using the
photon beam at MAMI \cite{Pfeiffer04}. It has, however, been
pointed out\cite{Hanhart05} that the data of Ref.\cite{Pfeiffer04}
does not permit an unambiguous determination of the existence of a
${\rm{^3He}} \eta$-bound state. The suggestion that ${\rm{^3He}} \eta$ is not
bound is also supported by the theoretical studies of Refs.\cite{Sofianos97, Haider_Liu02}.

Two different methods have been applied in the search for $\eta$ bound states. One is the study of $\eta$ production on nuclei and extraction of the $\eta$ nucleus scattering length. The other is the direct production of the $\eta$ meson in a bound state. This state is measured via missing mass technique. The GEM collaboration has used both methods, as will be shown in this paper.

\section{Searches via two body final state interaction}

According to the Watson-Migdal theory\cite{Watson52, Migdal55}, when
there is a weak transition to a system where there is a strong final
state interaction (\emph{FSI}), one can factorize the $s$-wave
reaction amplitude, $f_s$, near threshold in the form
\begin{equation}\label{equ:FSI}
f_s=\frac{f_B}{\frac{1}{a}+\frac{r_0}{2}p^2-iap}
\end{equation}
where $p$ is the $\eta$ c.m. momentum, $a$ the complex scattering length and $r_0$ the effective range. The unperturbed
production amplitude $f_B$ is assumed to be slowly varying and is
often taken to be constant in the near-threshold region.

Unitarity demands that the imaginary part of the scattering length be
positive, \emph{i.e.}, $ a_i>0$. In addition, to have binding, there
must be a pole in the negative energy half-plane, which requires
that\cite{Haider_Liu02}
\begin{equation}\label{equ:Cond2}
\left.{|a_i|}\right/{|a_r|}<1\,.
\end{equation}
Finally, in order that the pole lie on the bound- rather than the
virtual-state plane, one needs also $a_r<0$.

Recently, two different experiments at COSY J\"{u}lich measured $\eta$ production in $pd\to {\eta\rm{^3He}}$ reactions very close to threshold with an extremely high precision of the data\cite{Smyrski07, Mersmann07}.
The latter authors folded out the experimental resolution and got
$a_{{}\rm{^3He}\eta }  = \left[ { \pm \left( {10.7 \pm 0.8_{ - 0.5}^{ + 0.1} } \right) + i \cdot \left( {1.5 \pm 2.6_{ - 0.9}^{ + 1.0} } \right)} \right]{\rm{fm}}$
and
$r_0  = \left[ {\left( {1.9 \pm 0.1} \right) + i \cdot \left( {2.1 \pm 0.2_{ - 0.0}^{ + 0.2} } \right)} \right]{\rm{fm}}$
for the effective range. Since the data are not sensitive to the sign of the real part of the scattering length, the quest for a bound state or an unbound pole can not be answered. Its value is
\begin{equation}
|Q_{\rm{^3He}\eta } | \approx 0.30\,\,{\rm{MeV}}.
\end{equation}
From the model calculations it is known that binding is more probable for heavier nuclei than for lighter nuclei. We therefore can expect the relation
\begin{equation}\label{equ:Relation}
|Q_{\rm{^3He}\eta } |<|Q_{^4He\eta } |
\end{equation}
to hold. In the following we study of the FSI of the $\eta$ ${^4He}$ system. This is produced in the reaction
\begin{equation}
dd\to \eta\alpha .
\end{equation}
The existing data before the GEM measurement were not sufficient to extract the s-wave contribution of the cross section. In order to do so GEM made use of a tensor polarized deuteron beam\cite{Budzanowski09b}. The beam momentum of 2385.5~MeV/c
corresponds to an excess energy of $Q=16.6$~MeV for this reaction
when an $\eta$-meson mass of $m_{\eta}=547.7$~MeV/$c^2$ is
used~\cite{PDG}. The Big Karl magnetic spectrograph~\cite{drochner98,
Bojowald02} employed for this study is equipped with two sets of
multi-wire drift chambers (MWDC) for position measurement and thus
track reconstruction. Two layers of scintillating hodoscopes, 4.5~m
apart, led to a more accurate time-of-flight measurement than
previously achieved with Big Karl. They also provided the energy-loss
information necessary for particle identification. More information on the experiment are given in Ref.\cite{Budzanowski09b}.

In a first step the unpolarized cross section was measured.
\begin{figure}\begin{center}
\includegraphics[width=0.48\textwidth]{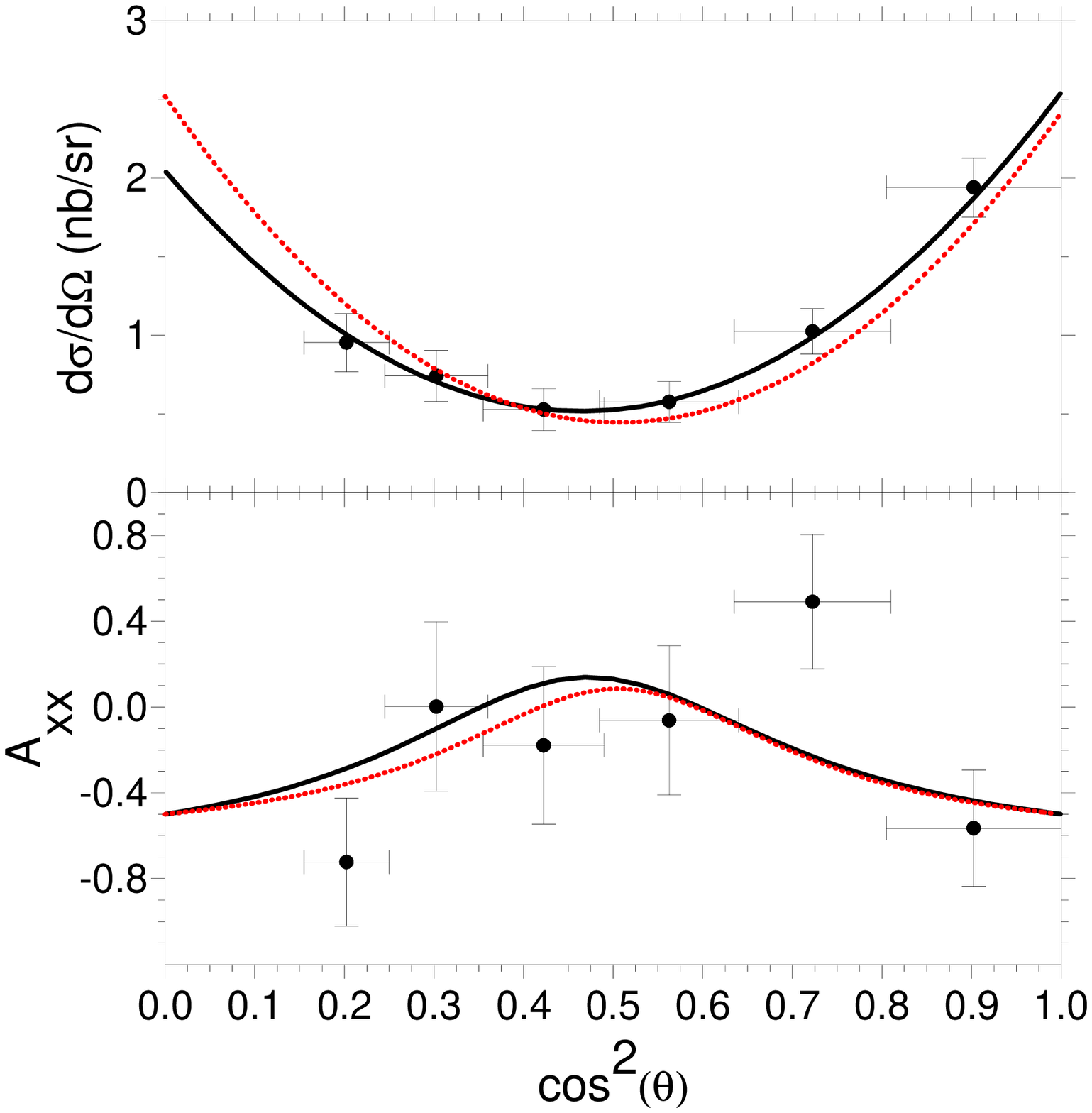}
\includegraphics[width=0.48\textwidth]{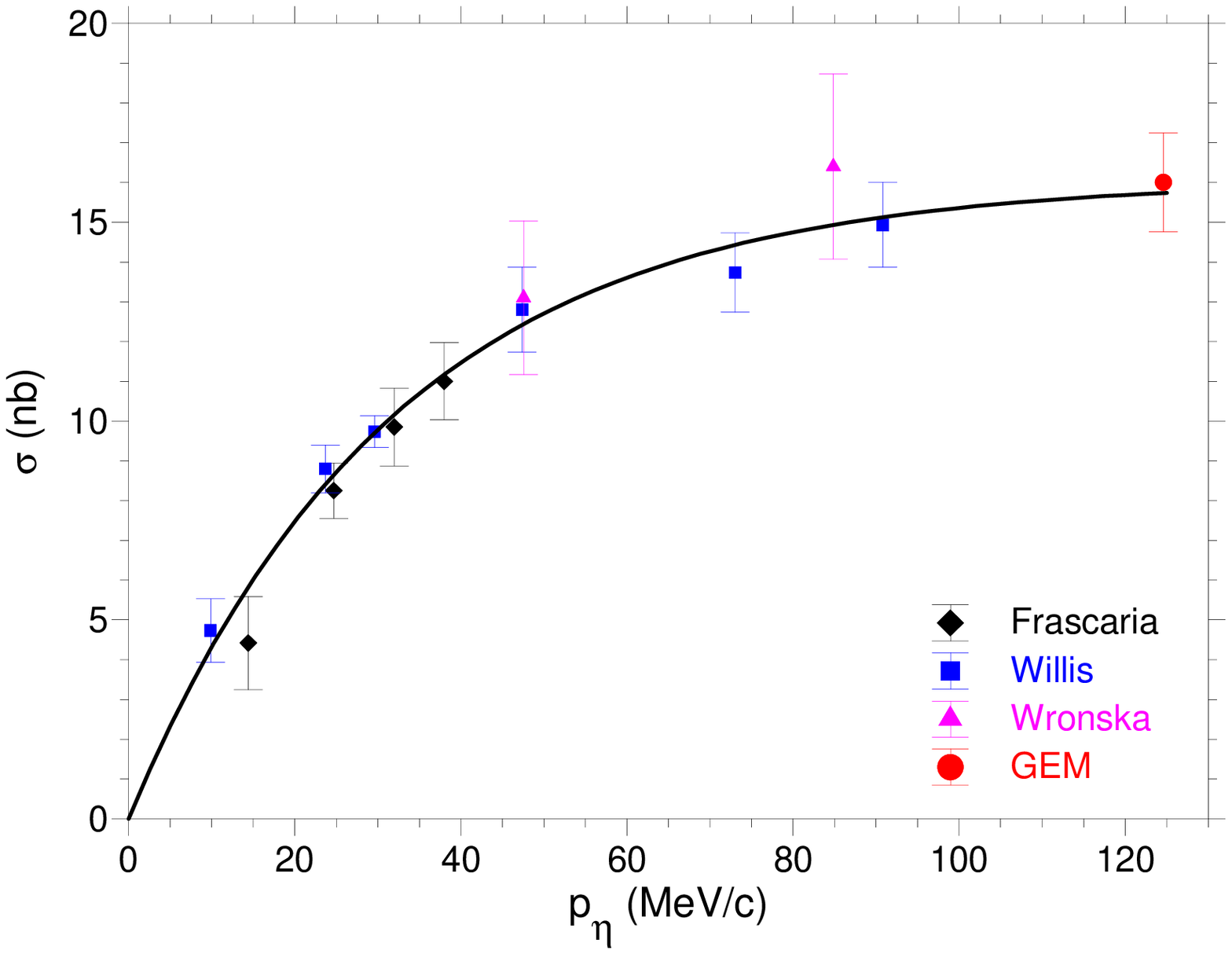}
\caption{\label{Fig:Partial_wave_fit}Left upper panel: Differential cross section for the
$dd\to\alpha\,\eta$ reaction. Left lower panel: Analyzing power $A_{xx}$.
The solid curves represent a fit with four partial waves; the dotted
curves with invariant amplitudes. Right: Excitation function for the total cross section for the
$dd\to \alpha\,\eta$ reaction.  Only statistical errors are shown. }
\end{center}\end{figure}
The obtained angular distribution is shown in Fig. \ref{Fig:Partial_wave_fit}. It can be fitted by Legendre polynomials. Because of the symmetric entrance channel only even polynomials contribute. Three parameters are sufficient to reproduce the data.
This indicates that there must be at least $d$-wave contributions.
This is to be contrasted to the lower energy ANKE
results~\cite{Wronska05}, where $2\ell_{max}=2$ suffices. From the fit total cross section of $\sigma = 16.0\pm0.4$ nb, where uncertainties in the target thickness, incident flux, and acceptance introduce an additional systematic error of $\pm 1.6$~nb.
This value is shown in Fig. \ref{Fig:Partial_wave_fit} together with the world data \cite{Wronska05}, \cite{Frascaria94}, and \cite{Willis97}. It seems that the cross sections start to saturate for $\eta$ momenta above 80 MeV/c.

The next step is to extract the s-wave part of the total cross section. For this task the knowledge of polarization observables is necessary. First the polarization of the beam $p_{zz}$ was measured by measuring elastic backward scattering of the deuterons on protons. From the known analyzing power $A_{yy}$ the beam polarization is obtained. In the measurement of analyzing powers one usually compares polarized and unpolarized cross sections. This introduces ambiguities because of different luminosity measurements. In order to avoid this we applied another method. We integrated the cross section over intervals in the polar angle, where on one hand we have full geometrical acceptance of the apparatus and $A_{yy}$  practically vanishes. We measure then only $A_{xx}$. the cross section integrated over these intervals of
azimuthal angle becomes simply
\begin{gather}
\nonumber I=\int_{(\pi-1)/2}^{(\pi+1)/2}\left(\frac{d(\theta,\phi)}{d\Omega}\right)_{\!\text{pol}}
d\phi =\\ \left(\frac{d{\sigma}}{d\Omega}
(\theta)\right)_{\!\text{unpol}} \left[1 + 0.46\,p_{zz}A_{xx}(\theta)
\right], \label{dsigma_rho_Axx}
\end{gather}
where the unpolarized cross section is integrated over the same
$\phi$ range.

Carrying out this procedure for the two polarization states, we find
that
\begin{gather}
\Delta = \frac{I^+-I_1^-}{I^++I^-} =
\frac{0.23\,A_{xx}\left(p_{zz}^+-p_{zz}^-\right)}{
1+0.23\,A_{xx}\left(p_{zz}^++p_{zz}^-\right)}\cdot
\end{gather}
and hence
\begin{equation}
A_{xx}=02.44\Delta .
\end{equation}
The angular distribution thus obtained is also shown in Fig. \ref{Fig:Partial_wave_fit}. From both angular distributions we extract partial waves, one s-wave, one p-wave and two d-waves. This yields seven parameters to be fitted. However, strong correlations were found. An alternative method is to employ the spin structure of the matrix element. Due to the symmetry in the entrance channel three independent scalar amplitudes are necessary to describe the spin dependence of the reaction: $A, B, C$.  $B$ and $C$ have no angular dependence and for $A$ we assume the expansion $A=A_0+A_2P_2(\cos \theta)$. Fortunately the dependencies can be decoupled
\begin{eqnarray}
\left(1 - A_{xx}\right)\frac{d\sigma}{d\Omega} &=&
\frac{p_{\eta}}{p_d}\left(|A_0|^2
+2\emph{Re}(A_0A_2^*)P_2(\cos\theta)+|A_2|^2\left(P_2(\cos\theta)\right)^2\right)\,,
\label{comb1}\\
\left(1 + 2 A_{xx}\right)\frac{d\sigma}{d\Omega}
&=&2\frac{p_{\eta}}{p_d}\left(|B|^2\sin^2\theta\cos^2\theta+|C|^2\sin^2\theta\right).
\label{comb2}
\end{eqnarray}

\noindent We now extract the magnitude of the $s$-wave amplitude $|a_0|$.
From this we obtain a spin-averaged square of the $s$-wave amplitude,
$|f_s|^2$ through
\begin{equation}\label{Eq:f_s}
\frac{d\sigma_s}{d\Omega}=\frac{p_\eta}{p_d}|f_s|^2 =
\frac{2p_\eta}{3p_d}|A_0|^2= \frac{1}{27} \frac{1}{4\pi}|a_0|^2 .
\end{equation}
Wro\'{n}ska et al.\cite{Wronska05} could not distinguish whether the angular distribution is due to a $s-p$ interference or a $d$-wave. This puzzle can be solved by our result here. The Willis et al.\cite{Willis97} data for s-wave were extracted from the measurement by just dividing by $4\pi$. This is, however, not justified in the region of their highest point. Assuming a dependence of the $d$-waves with $p_\eta^2$ we can extract a more correct value for the s-wave strength. From all measurements we then obtain a scattering length of
\begin{equation}
a_{{}^4He\eta }  = \left[ { \pm \left( {3.1 \pm 0.5} \right) + i \cdot \left( {0.0 \pm 0.5} \right)} \right]{\rm{fm}}.
\end{equation}
This result can be converted into a pole position
\begin{equation}
|Q_{^4He\eta }|\approx 4 \text{MeV}.
\end{equation}
Since the present results fulfill the conditions for a bound state - except for the sign of the real part of the scattering length - and the relation \ref{equ:Relation}, it may well be that $\eta-\alpha$ binding is observed.

For an even heavier system we may expect stronger binding. GEM has therefore studied $\eta$ production in an almost exclusive reaction $p+^6Li\to\eta^7$Be, by measuring the recoiling Be ions. Since the energy of the ions is rather small, they are strongly ionizing particles and new detectors have been applied. Details are given in \cite{PLI6}. Since all states of $^7$Be above 1.59 MeV are particle unstable, only the two lowest states with $L=1$ contribute.
\begin{figure}[h]
\begin{center}
\includegraphics[width=0.5\textwidth]{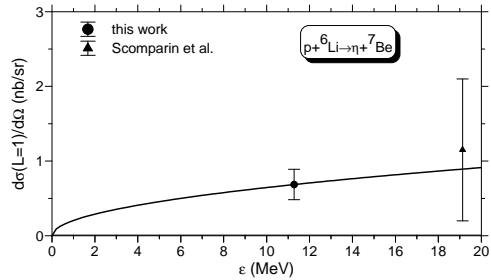}
\caption{Differential cross sections for the indicated reaction with $^7$Be in its ground state or first excited state. The solid curve is phase space behavior fitted to the present measurement.}
\label{Fig:pli}
\end{center}
\end{figure}
In Fig. \ref{Fig:pli} the new data point at 11.3 MeV excess energy is shown together with a point earlier obtained by \cite{Scomparin93} and corrected for $L=3$ contributions. The data can obviously be accounted for by phase space. More data closer to threshold are necessary to see FSI effects.

\section{Production of $\eta$ bound nuclear states}
The following method was successfully employed in the production of Hypernuclei and pionic atoms. A nucleon is replaced by a hyperon or a pion. Maximal cross section is obtained when the momentum transfer from the projectile to the particle to be implanted is minimized, which means that the produced particle is almost at rest in the laboratory system. In pionic atom studies a proton transfer via the $(d,{\rm{^3He}})$ reaction was successfully applied\cite{Yamazaki96}. However, although this reaction has reasonably large cross section it also has the disadvantage of a huge background due to the break-up  of the deuterons. The resulting protons have the beam velocity and thus the same magnetic rigidity as the ${\rm{^3He}}$ particles. This problem can be overcome by making use of a two-nucleon transfer reaction like $(p,{\rm{^3He}})$ but at the expense of a much smaller cross section. The choice of an odd-odd target nucleus would be ideal in order to avoid nuclear excitations as much as possible. Since there exist no solid material with these properties $^{27}$Al was chosen as a compromise. The experiment was done by making use of the following reaction chain:
\begin{equation}\label{equ:step1}
p+^{27}\!\text{Al}\to \rm{^3He}+X
\end{equation}
where the $\rm{^3He}$ carries the beam momentum away. The unobserved system $X$ is therefore at rest. One of the possibilities is $X=^{25}Mg+\eta$. Since the $\eta$ is also at rest it can undergo a second chain of reactions
\begin{equation}\label{equ:step2}
\eta + N\leftrightarrows N^*\to N\pi.
\end{equation}
The final fate of that chain is the decay to a pion and a nucleon. In the case of a neutral resonance the final state can be $p+\pi^-$, which have to be emitted almost back to back, if we ignore Fermi motion. The $\rm{^3He}$ is detected with the magnetic spectrograph Big Karl\cite{drochner98,
Bojowald02}. For the detection of the two decay particles a dedicated detector ENSTAR was built. It has cylindrical shape around the target and consists of three layers of scintillating material. The layers were subdivided into pieces making polar and azimuthal angle measurements possible. The read-out of the pieces was performed with scintillating fibres\cite{Betigeri07}.
\begin{figure}[!h]
\begin{center}
\includegraphics[width=0.5\textwidth]{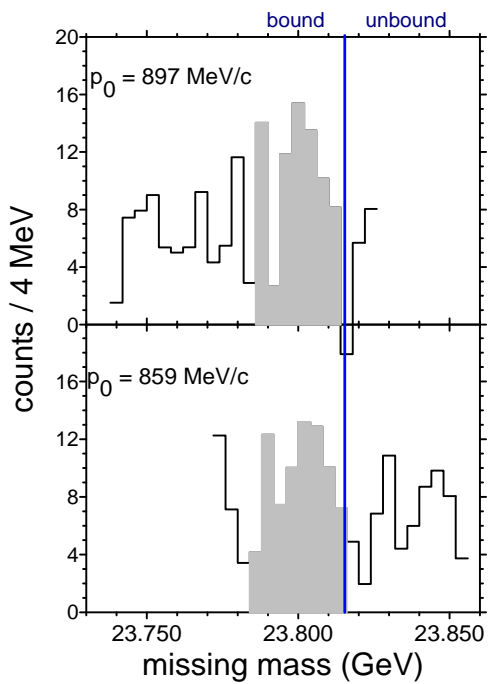}
\caption{Missing mass spectra for the reaction $p+^{27}$Al$\to ^3$He$+p+\pi^-+X$ at recoil free condition for an $\eta$.}
\label{Fig:two_settings}
\end{center}
\end{figure}

The experiment was performed at a beam momentum of 1745 MeV/c. At this momentum final states with small binding energies can be produced with small momentum transfer. States with binding energies from 0 MeV to -30 MeV will have momenta below 30 MeV/c.

Two settings of the spectrograph were used with a large overlap. Excellent particle identification in Big Karl was achieved by measuring energy loss in the start detector of the TOF facility in the focal plane and the TOF. The identification of a back to back proton and pion with momenta corresponding to a $N^+$(1535) resonance is mandatory to see structure in the missing mass spectra. The spectra for the two settings are shown in Fig.~\ref{Fig:two_settings}. The shaded area may correspond to an $\eta$ bound state. The data in the lower panel contain the unbound region with quasi-free $\eta$ production. More details can be found in \cite{Budzanowski10}.

The two spectra were then combined. Counts were transformed into differential cross sections by making use of the integrated luminosity and detector acceptance as well as its efficiency.
\begin{figure}[h]\begin{center}
\includegraphics[width=8cm]{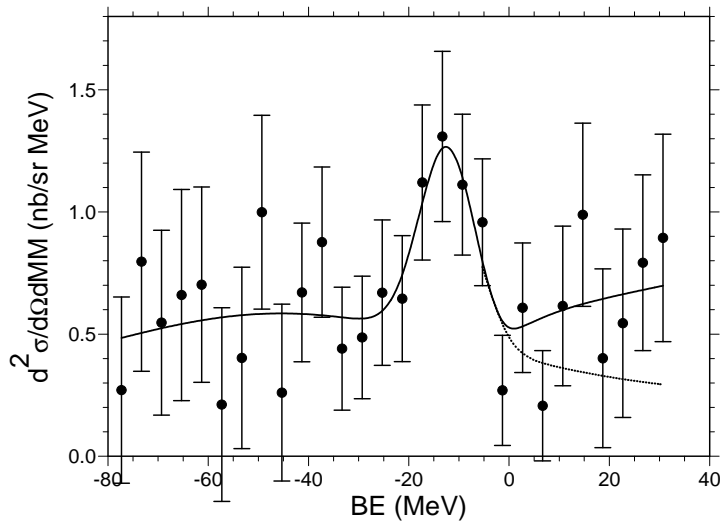}
\caption{\label{Fig:Binding-spectra}Missing mass spectrum converted to binding energy $BE$ of a bound system $^{25}Mg\otimes\eta$ as measured with the magnetic spectrograph. All cuts and background subtraction have been applied. The solid curve is a fit with a constant background, two Gaussians and a phase space behavior for the unbound system. Dotted curve: same as solid curve but without phase space contribution. }
\end{center}\end{figure}
These cross sections are shown in Figure \ref{Fig:Binding-spectra}. here we have converted the missing mass scale into binding energy $BE$. The data show an enhancement around  $BE\approx
-13$~MeV. The significance of this structure is
extracted according to the two methods given in \cite{FST_statistics}. At first, we test the hypothesis of peak
structure being a fluctuation of background, i.e. the origin of
the background is taken to be independent of the signal.
The background outside the peak region, for simplicity approximated by a constant, was found to be 5.8$\pm$0.64. The significance is then given by $(N-BG)/\sqrt{BG+\sigma_{BG}}$
where $N$ is the total counts in the region of interest, $BG$ is the
total background in this region as determined from the fit to the
outside region and $\sigma_{BG}$ is error in the estimation of
background value as taken from the fit. This yields a value of
significance which is 5.3$\sigma$. Here we have assumed Gaussian errors. For the assumption of Poisson errors the background is 6.2$\pm$1.0. This larger value is typical for Poisson distribution and hence the significance reduces to 4.9$\sigma$. Finally a Gaussian on top of the background was fitted to the whole data set. This yielded for the case of Poisson statistics 6.4$\pm$0.96 for the background, 8.3$\pm$3.6 for the amplitude, -12.0$\pm$2.2~MeV for the centroid and 4.7$\pm$1.7~MeV for the width.
In the second method, the statistical significance is extracted by assuming the background events as well as the peak events on top of the background being Poisson distributed. Again a constant background and a Gaussian was assumed.  In this way, we obtain a value of 6.20$\sigma$ for the
significance, assuming a simultaneous determination of amplitude,
centroid and width of the signal. The fit gives for the centroid $-13.13\pm 1.64$~MeV and
for the width $4.35\pm 1.27$~MeV. These results compare favorably with those from the first method. We, therefore, consider the present experimental results to provide a strong hint of a nuclear $\eta$ bound state.

In the following we take a different view on the data than in \cite{Budzanowski10}. It may well be that in addition to a $1s$ state a $1p$ state is bound if the scattering length $a_{\eta N}$ is large. The former state is stronger bound than the later and it is expected to have a larger width than the later \cite{Haider_Liu02}. We therefore fit two Gaussians on a constant background to the data. The quasi-free region is represented by phase space behavior.
\begin{table}[h]
  \centering
  \caption{Fitted parameters of two Gaussians to the data and their hypothetical assignment.}\label{Tab:1}
  \begin{tabular}{cccc}
  \hline
state & amplitude (nb/sr MeV) & centroid (MeV) & width $\sigma$ (MeV) \\ \hline
$1s$ & $0.33\pm0.08$ & $-45.4\pm19.0$ & $37.7\pm26.4$ \\
$1p$ & $0.87\pm0.21  $ & $-12.3\pm1.2$ & $5.1\pm1.6$ \\ \hline
\end{tabular}
\end{table}
The constant background was found to be 0.25 nb/sr MeV and the fit parameters of the Gaussians are given in Table \ref{Tab:1}. The resulting curve is also shown in Fig. \ref{Fig:Binding-spectra}. The hypothetical $1s$ state has a very large width and reaches into the unbound region. However, the evidence for this peak is less than the one for the peak at 12 MeV. Due to the low statistics of the data, which is a result of the small cross section, the uncertainties in the fitted parameters are large.

\section{Summary}
We have measured $\eta$ production in two body final state by $dd\to\eta\alpha$ and $p^6$Li$\to\eta^7$Be reactions close to threshold. For both reactions angular distributions were obtained which allow to extract total cross sections. Because of making use of a tensor polarized deuteron beam an average matrix element for the s-wave could be extracted and thus allows the determination of the scattering length. From the data analysis it is found that $p$-waves are absent. The data situation in the case of the beryllium final state is insufficient to draw conclusions. We have furthermore measured $p^{27}$Al$\to^3$He$p\pi^-$X at kinematical conditions where an $\eta$ in the elementary reaction $pd\to\eta^3$He is at rest. The missing mass spectrum shows a significant peak. There is evidence that the $\eta$ may be bound in a $1s$ and a $1p$ state. However, better statistics data are mandatory in order to identify the nature of the missing mass spectrum.

{The author is grateful to the members of the GEM collaboration.}

%


\end{document}